\documentstyle[pra,aps,amssymb,twocolumn,epsfig]{revtex}

\begin{document}

\draft

\twocolumn[\hsize\textwidth\columnwidth\hsize\csname
@twocolumnfalse\endcsname
\title{Josephson tunneling between weakly interacting Bose-Einstein 
condensates}
\author{F. Meier and W. Zwerger}
\address{ Sektion Physik, Universit\"at M\"unchen, Theresienstrasse 
37, 
D-80333 M\"unchen, Germany }
\date{\today}

\maketitle

\begin{abstract}
Based on a tunneling Hamiltonian description, we calculate the 
Josephson, normal and interference currents between two Bose-Einstein 
condensates described by the Bogoliubov theory. The dominant 
Josephson term is of first order in the tunneling with a critical 
current density proportional to the ground state pressure. In 
contrast to superconductors, the normal current remains finite at zero 
temperature. We discuss the dynamics of the relative phase in a
semiclassical approximation derived from an exact functional integral 
approach, which includes the interaction effects at fixed total particle 
number. It is shown that the normal current leads to a 
damping of the Josephson oscillations and, at long times,  
eliminates the macroscopic quantum self trapping predicted
by Smerzi et.al.
Finally we give estimates for an experimental 
realization of Josephson tunneling in cold atomic gases, which 
indicate that coherent transfer of atoms might be realized with 
a $^{23}$Na condensate.
\end{abstract}

\vskip2pc]

\section{Introduction}

The realization of Bose-Einstein condensation (BEC) in atomic vapors 
\cite{and95,dav95,bra97} provides an example of superfluidity, to 
which the approximation of a weakly interacting Bose gas 
applies quantitatively. Due to the
low densities $n\approx 10^{14}$ cm$^{-3}$ and with typical
scattering lengths $a\approx 5$nm, the gas parameter $na^3$
is of order $10^{-5}$, i.e. well in the range of applicability
of the Bogoliubov theory. Although the interactions still
entail strong quantitative changes compared to an ideal
gas picture~\cite{baym96}, the systems are rather well described by
the weak coupling Gross-Pitaevski equation (GPE) and
its small fluctuations, the Bogoliubov equations~\cite{str98}.
Many of the phenomena predicted 
within this framework like sound modes with a 
linear spectrum have been established experimentally 
\cite{and97}. More recently genuine superfluid properties like 
quantized vortices \cite{matthews99,dalibard00} or the 
existence of a critical velocity
\cite{raman99} have also been observed.

The Josephson effect \cite{jos62,jos69} is one of the prime examples of 
macroscopic quantum effects, displaying directly the broken symmetry
associated with the relative phase of two weakly coupled condensates
\cite{anderson66}. In superconductors it is a rather standard 
phenomenon,
in contrast to Bose condensed systems, where it has first 
been observed by Avenel and Varoquaux in superfluid $^4$He 
\cite{ave85}. For dilute atomic gases, coherent oscillations  
have been seen in driven two 
component BECs \cite{hal98} and in vertical arrays of 
traps of an optical lattice \cite{and98}. However these are
Rabi- or Wannier-Stark-type oscillations
and should not be confused with a genuine ac-Josephson effect as
discussed here. 
Theoretically, the Josephson effect in this context has been 
investigated by Smerzi {\it et al.} \cite{sme97}, who focussed on the 
nonlinear dynamics of the relative phase and population difference.
Zapata {\it et al.}\ \cite{zap98} 
calculated the Josephson coupling energy due to 
condensate-condensate tunneling for weakly coupled harmonic wells 
and gave estimates for the normal currents due to the 
thermal cloud. These studies were essentially limited to a mean field
like Gross-Pitaevski description, neglecting  
noncondensate contributions. More recently Villain and Lewenstein
\cite{villain99} showed that particles out of the condensate lead to
a damping of the relative phase in a two component condensate
with a Josephson like coupling induced by the external driving field
\cite{hal98}.  

Our aim in this work is to provide a microscopic calculation of the
Josephson effect between two weakly interacting BECs separated by
a tunnel barrier~\cite{fz99}. Using an idealized model, we give quantitative 
results for both the superfluid and normal currents per area 
which only depend on the microscopic parameters of the condensate and 
the barrier transmission amplitude. Moreover, a complete theory is
given for the interaction effects arising in coupled condensates 
with a fixed total particle number in the limit, where the number
of exchanged particles is small. The 
associated dynamics is discussed within a semiclassical 
approximation.
Using realistic parameters in dilute atomic gases, the  
effects due to a fixed number of particles 
are appreciable, but do not destroy the qualitative structure
of oscillating Josephson currents for condensates at different 
chemical potentials. Our estimates indicate that the Josephson effect
might be observable with currently available techniques for $^{23}$Na
or lighter atoms.

The outline of the paper is as follows: In section 2, we introduce 
the tunneling Hamiltonian model, including a discussion of 
the relevant states in condensates
with a fixed total particle number and the general form
of the Josephson coupling energy. This energy and the
associated nondissipative current is  
calculated up to second order in the tunneling matrix 
elements in section 3. For a finite, given difference in the
chemical potentials we determine in section 4 the complete 
particle current in terms of the normal and anomalous spectral 
functions of the individual condensates. In section 5, we develop an exact 
functional integral representation of the reduced quantum mechanics of 
weakly coupled Bose condensates which incorporates the effect 
of interactions at fixed total
particle number. In section 6, explicit results are given
for an idealized geometry, determining
the physically relevant currents per area from
microscopic parameters which characterize the separate condensates. 
The resulting 
semiclassical dynamics of the relative phase, including `charging' 
effects and 
dissipation, is solved numerically in section 7 for parameters, which 
are 
realistic for current condensates of dilute atomic gases. Moreover, 
we test the 
tunneling Hamiltonian Ansatz by comparing it with exact numerical 
results for the one-dimensional GPE with a barrier. 
Conclusions and 
open questions are discussed in section 8.

\section{The tunneling Hamiltonian}

Conceptually, the most obvious way to realize a Josephson geometry  
for BECs with current experimental setups, is to split a single 
condensate in a long trap into two separate parts by
a narrow light sheet produced by a blue detuned laser. 
The repulsive
potential due to the ac-Stark shift is proportional to the
laser intensity and thus the height and width 
of the barrier may be varied in a considerable range. 
In the limit of a strong barrier one has to zeroth«
order two completely separate condensates $a$ and $b$ with 
Hamiltonians $\hat{H}_{a/b}$. Provided that 
the coupling energy $E_{J}$ due to the transfer of particles 
across the barrier is 
small compared to the ground state energies of $\hat{H}_{a/b}$, 
which are of order $\mu N$, 
the coupled system may then be approximated by a 
tunneling or transfer Hamiltonian \cite{duk69}
\begin{equation}
\hat{H} = \hat{H}_a + \hat{H}_b + \hat{H}_T
\label{eq:th1}
\end{equation}
in which the contribution $\hat{H}_T = \hat{\Lambda} + 
\hat{\Lambda}^{\dagger}$ describing the transfer
of particles between $a$ and $b$ can be treated 
perturbatively. Formally, introducing a set of 
orthonormal eigenstates
$\{ | l \rangle \}$ and $\{ | r \rangle \}$ 
which decay exponentially in the barrier where $a$ and $b$
overlap, an instantaneous transfer is described by a term
\begin{equation}
\hat{\Lambda} = -\sum_{l,r} t_{lr} \hat{a}^{\dagger}_l \hat{a}_r
\label{eq:th2}
\end{equation}
with tunneling amplitudes $t_{lr}$, which may be 
expressed in terms of current matrix elements between
the left and right eigenstates~\cite{duk69} (see section
6 below). The signs here have been choosen in such a way
that for positive, real ground state wave functions in $a$
and $b$, the associated tunneling amplitude is positive,
giving rise to a lowering of the ground state energy in the
coupled system, as expected in a standard double well
situation. While a rigorous
derivation of the transfer Hamiltonian is rather
difficult~\cite{duk69}, it is expected to be valid for 
high and narrow barriers, for which the
mean field interaction of the condensate is
negligible within the barrier. This is in fact
confirmed through numerical calculations in an exactly
soluble one-dimensional geometry in section 7. Note that 
the states $\{ | l \rangle \}$ and $\{ | r 
\rangle \}$ introduced in Eq.~(\ref{eq:th2}) are not 
mutually orthogonal, and thus $\hat{H}_{a}$ and $\hat{H}_{b}$ do
not commute beyond zeroth order in $\hat{H}_{T}$~\cite{pra63}.
Moreover, the creation and annihilation operators in (2)
refer to the transfer of atoms, not that of quasiparticles,
which are the proper excitations only within the
individual condensates.

As pointed out by Ferrell and Prange~\cite{fp63}, the essence
of the Josephson effect is that in the absence of any drop
$\Delta\mu$ in the chemical potential between both sides,
no energy is required to transfer a condensate atom across
the barrier (note that a finite value of $\Delta\mu$ may
be present either {\it ex}ternally e.g. by a drop in the
gravitational potential as in the experiment by Anderson
and Kasevich~\cite{and98}, or may arise {\it in}ternally
at finite transfer currents through `charging' effects
discussed in section 5 below). Thus the states
$|\nu\rangle =|\bar{N}_{a}+\nu\, ,\bar{N}_{b}-\nu\rangle$ in which 
an arbitrary integer number $\nu$ of condensate atoms have 
tunneled from right to left, are degenerate. The ground
state of the coupled system is therefore not an eigenstate
$|\nu\rangle$ with a definite relative particle number, but
rather a coherent superposition
\begin{equation}
|\varphi\rangle=\sum_{\nu}\, c_{\nu}e^{i\nu\varphi}|\nu\rangle
\label{eq:th3}
\end{equation}
with a definite relative phase $\varphi=\varphi_{a}-\varphi_{b}$.
Here the $c_{\nu}$ are real coefficients obeying
$\sum\, c_{\nu}^2=1$ which are constant around
$\nu=0$, decaying to zero only at $|\nu|$ of order
$\sqrt{N_{a/b}}$. In terms of the eigenstates
$|\varphi_{a},\varphi_{b}\rangle$ in which each of the condensates
has a definite overall phase $\varphi_{a,b}$, the state
$|\varphi\rangle$ can be expressed in the form
\begin{equation}
| \varphi \rangle \propto \int_0^{2 \pi} \frac{d \varphi_b}{2 \pi} 
e^{- i (N_a + 
N_b) \varphi_b} |\varphi_b + \varphi ,\varphi_b 
\rangle ,
\label{eq:th4}
\end{equation}
using the standard relation between states with a fixed
phase or a fixed particle number~\cite{anderson66}.
The state with a definite {\it relative} phase is thus
a projection of states $|\varphi_{a},\varphi_{b}\rangle$ with a 
broken gauge symmetry in the individual condensates to
one with a definite total particle number $N_{a}+N_{b}$.

The delocalization in the relative particle number $\nu$
described by the coherent state $|\varphi\rangle$, gives rise to a 
lowering of the energy of the coupled condensates in the 
same manner, in which a Bloch state gains kinetic energy by
spreading over many sites in a periodic potential with
localized Wannier orbitals labeled by $\nu$. In a time
reversal invariant situation, this energy (or free
energy at $T\not= 0$) must be even and periodic in
$\varphi\to\varphi +2\pi$. Quite generally it can thus be
expanded in a Fourier series of the form~\cite{bloch70}
\begin{equation}
E(\varphi)=\sum_{n=0}^{\infty}\, E_{n}\cos{n\varphi}.
\label{eq:th5}
\end{equation}
Since $\nu$ and $\hbar\varphi$ are canonically conjugate
variables, the nondissipative current 
\begin{equation}
I(\varphi)=-\frac{d\nu}{dt}=-\frac{\partial E(\varphi)}{\partial\,
\hbar\varphi}=\sum_{n=1}^{\infty}\, I_{n}\sin{n\varphi}
\label{eq:th6}
\end{equation}
can be obtained from the knowledge of the phase dependent part
of the coupling energy between the two condensates. Within a
tunneling Hamiltonian description, the coefficients $E_{n}=
\hbar I_{n}/n$ are of order $|\hat H_{T}|^{n}$. It is thus
usually sufficient to keep only the lowest nonvanishing term,
conventionally denoted by $-E_{J}$, which is negative in the
standard situation, where the lowest energy state is that with
a vanishing relative phase $\varphi=0$. As will be shown below,
for Bose condensates the Josephson coupling arises already in first 
order in the tunneling amplitudes,
\begin{equation}
<\varphi|\hat H_{T}|\varphi>=-E_{J}\,\cos{\varphi}
\label{eq:th7}
\end{equation}
with a positive Josephson coupling energy
$E_{J}$, which immediately determines the associated
critical current $I_{c}=E_{J}/\hbar$. For
superconductors, in turn, where the Josephson effect is associated
with the coherent tunneling of {\it pairs}, the first order
term vanishes and $E_{J}$ is proportional to the average
of $|t_{lr}|^2$ at the Fermi energy. The latter being a direct
measure also of the {\it normal} currents, one obtains the well 
known relation 
\begin{equation}
2E_{J}^{BCS}=\Delta\tanh{\beta\Delta/2}\cdot hG_{n}/{4e^2}
\label{eq:th8}
\end{equation}
connecting the BCS Josephson coupling to the 
superconducting gap $\Delta$ and the
dimensionless normal state conductance $G_{n}$
~\cite{amb63}.

Since tunneling is a small perturbation, the fact that 
there is a Josephson effect already in {\it first} order for
a Bose condensed system, might seem to make higher order
calculations unnecessary in this case. In fact, there are
two reasons why this is not so: First of all, it turns out
that realistic barriers which allow to observe a 
Josephson effect in coupled BECs have to be small enough,
that higher order effects are nonnegligible. More
importantly, though, it is only at second order that
dissipative effects appear through normal currents which,
as we will see, can have strong effects even if they are 
small.
 
\section{The phase dependent coupling energy}

In order to
calculate the change in energy due to tunneling, we employ
perturbation theory up to second order in the transfer term
$\hat H_{T}$. To be specific, we assume that the condensate
size is large compared with the coherence length 
$\xi=(8\pi na)^{-1/2}$ over
which the condensate wave function varies~\cite{baym96}.
Under this condition, which is well realized in present
gaseous BECs, the eigenstates $|l \rangle$ and $|r \rangle$ introduced
above for a weakly interacting gas are given by
\begin{itemize}
\item the condensate ground states $\phi_{a/b}(x/y)$
which follow from a solution of the Gross-Pitaevski
equation in the given confining potential, and
\item a set of excited states orthogonal to
$\phi_{a/b}(x/y)$, which may be labeled by nonzero
wavenumbers $k$ or $q$ for condensates $a$ and $b$ 
respectively. These states are obtained by solving the
standard Bogoliubov-equations in a local density or -
equivalently - a semiclassical approximation,
resulting in wavefunctions and energies like that in
a homogeneous system with a local condensate density
$n_{c}(x)$~\cite{giorgini97}. 
\end{itemize}
The transfer operator
\begin{eqnarray}
\hat{\Lambda} & = & -t_{cc} \hat{a}^\dagger_{c,a} \hat{a}_{c,b} -
\sum_{k}\, t_{kc} \hat{a}^\dagger_{k} 
\hat{a}_{c,b}  \nonumber \\ && \hspace*{1.5cm} -
\sum_{q}\,
t_{cq} \hat{a}^\dagger_{c,a} \hat{a}_{q} -  
\sum_{k,q}\, t_{kq} 
\hat{a}_{k}^\dagger \hat{a}_{q}
\label{eq:ce1}
\end{eqnarray}
is thus split into contributions describing 
condensate to condensate (c-c), condensate to noncondensate
(c-nc) and noncondensate to noncondensate (nc-nc) 
tunneling. To first order in $\hat H_{T}$, only the c-c
term contributes and gives rise to an energy gain
of precisely the form (7),
with a positive Josephson coupling energy
$E_{J}^{BEC}=2t_{cc}(\bar N_{a}\bar N_{b})^{1/2}$, which favors
a fixed relative phase $\varphi=0$ in the ground state of
the coupled system. Here, as in the rest of our work, we
have assumed that $|\nu|$ remains much smaller than the 
average number ${\bar N}_{a/b}$ of condensate atoms in
each well. The Josephson coupling energy is thus
independent of the number of transferred atoms to lowest
order.

For the calculation of the second order energy shift
\begin{equation}
\Delta E^{(2)}=-\sum_{e}\,\frac{\vert\langle e|\hat H_{T}|0\rangle
\vert^2}{E_{e}-E_{0}}
\label{eq:ce2}
\end{equation}
involving all possible 
excited states $|e\rangle$, we follow a derivation given by
Ferrell for the analogous case of Josephson tunneling 
between two BCS superconductors÷\cite{ferrell88}. 
As is evident from (9), there are
three contributions to the transfer Hamiltonian involving
excited states: two c-nc terms and one nc-nc contribution.
Denoting time reversed states by $\bar k=-k$ and using the
time reversal invariance relation $t_{ck}=t_{\bar{k}c}$,
the contribution which describes tunneling between condensate
$b$ and noncondensate states in $a$ can be written as
\begin{equation}
\hat H_{T}^{c_{b}-{nc}_{a}}=-\sum_{k}\, t_{kc}\left(
\hat a_{k}^\dagger \hat a_{cb}+\hat a_{cb}^\dagger \hat a_{\bar{k}}
\right).
\label{eq:ce3}
\end{equation}
Starting in a state $|\nu\rangle$ with a definite number of
atoms in each condensate but no excitations, the relevant
excited states $|k,\nu\rangle$ associated with the first term
in (11) are those, in which a condensate atom from $b$ is
transferred to a quasiparticle $k$ in $a$. With the 
standard Bogoliubov transformation
\begin{equation}
\hat{a}_{k}^\dagger = u_k \hat{\alpha}_{k}^\dagger - v_k 
\hat{\alpha}_{ 
\bar{k}}
\label{eq:ce4}
\end{equation}
between the Boson or quasiparticle creation operators
$\hat a_{k}^\dagger$ or $\hat\alpha_{k}^\dagger$, the 
associated matrix element is 
\begin{equation}
\langle k,\nu |\hat H_{T}^{c_{b}-{nc}_{a}}|\nu\rangle 
=-t_{kc}u_{k}\sqrt{\bar N_{b}}\, ,
\label{eq:ce5}
\end{equation}
again assuming $|\nu |\ll\bar N_{b}$. Now precisely
the same excited state with energy $E_{k}$ (both $E_{k}$
and the particle and hole amplitudes $u_{k}$ and $v_{k}$
are positive and even in $k$, following standard notation
~\cite{fet71}) can be reached from the state $|\nu +2\rangle$
by converting two condensate atoms in $a$ to a pair
$(k,\bar k)$ and transferring one of the partners $\bar k$
to a condensate state in $b$, as indicated by the second
term in (9). The amplitude for this process is
\begin{equation}
\langle k,\nu |\hat H_{T}^{c_{b}-{nc}_{a}}|\nu +2\rangle 
=t_{kc}v_{k}\sqrt{\bar N_{b}}.
\label{eq:ce6}
\end{equation}
It differs from (13) by the replacement $u_{k}\to -v_{k}$
since the amplitude for creating a quasiparticle with
momentum $k$ by destroying an atom with momentum $-k$ is
$-v_{k}$. In more physical terms, the relative minus sign
between these amplitudes may be understood by noting, that
in the Bogoliubov theory (and in fact quite generally for
a Bose condensed system) there is a fixed relative phase
$\pi$ between the condensate and pairs $(k,\bar k)$ of
noncondensate atoms~\cite{noz83}. This minimizes the
repulsive interaction, giving rise to a gapless excitation 
spectrum in contrast to the result of a Hartree-Fock 
approximation. The phase locking between condensate and
noncondensate is also implicit in the form
\begin{equation}
|\varphi_{a}\rangle\propto\exp{\left( e^{i\varphi_{a}}\sqrt{N_{a}}
\hat a_{c,a}^{\dagger}-e^{2i\varphi_{a}}\sum_{k}
\frac{u_{k}}{v_{k}}\hat a_{k}^{\dagger}\hat a_{\bar k}^{\dagger}
\right)}
\label{eq:ce7}
\end{equation}
of the Bogoliubov ground state for a homogeneous system
with a well defined overall phase $\varphi_{a}$, as 
introduced in (4) above. Now, as emphasized before, the
ground state of the coupled condensates is one with a
definite relative phase $\varphi$, in which the amplitudes
(13) and (14) add coherently
\begin{equation}
\langle k,\nu |\hat H_{T}^{c_{b}-{nc}_{a}}|\varphi\rangle 
=-e^{i\nu\varphi}t_{kc}\sqrt{\bar N_{b}}\left( u_{k}-
e^{2i\varphi}v_{k}\right).
\label{eq:ce8}
\end{equation}
With $E_{k}$ as the relevant excitation energy, this gives 
rise to a phase dependent contribution
\begin{equation}
	\bar N_{b}\sum_{k}\,\frac{|t_{kc}|^2}{E_{k}}\,
	2u_{k}v_{k}\cdot\cos{2\varphi}
\label{eq:ce9}
\end{equation}
to the second order energy shift. Using $2u_{k}v_{k}=
\mu_{a}/E_{k}$ and exchanging the roles of condensates 
$a$ and $b$, the total c-nc contribution to the phase
dependent energy in second order is
\begin{equation}
\left( \bar N_{a}\mu_{b}
\sum_{q}\,\frac{|t_{cq}|^2}{E_{q}^2}\, +
\bar N_{b}\mu_{a}
\sum_{k}\,\frac{|t_{kc}|^2}{E_{k}^2}\right)\cdot\cos{2\varphi}.
\label{eq:ce10}
\end{equation}
Obviously this term, which is much larger than the nc-nc
contribution derived below, favors a relative phase 
$\varphi =\pi/2$ of the coupled condensates. In the weak 
tunneling regime considered here, however, it is always
the leading first order term (7) which dominates, leading 
to a ground state with $\varphi =0$. To obtain the nc-nc
contribution, we use again time reversal invariance to write
\begin{equation}
\hat H_{T}^{nc-nc}=-\sum_{k,q}\, t_{kq}\left(
\hat a_{k}^\dagger \hat a_{q}+\hat a_{\bar q}^\dagger \hat a_{\bar{k}}
\right)\, .
\label{eq:ce11}
\end{equation}
The relevant excited states $|k,\bar q,\nu\rangle$ now
have energy $E_{k}+E_{q}$ and are characterized by
$\bar N_{a}+\nu$ and $\bar N_{b}-\nu -2$ atoms in condensates
$a$ and $b$ plus one quasiparticle $k$ respectively $\bar q$  
on each side.  Corresponding to the two contributions in (19),
these states can be reached starting from either $|\nu\rangle$ or
$|\nu +2\rangle$. Adding the respective amplitudes in the
actual coherent ground state with a well defined relative 
phase, one obtains
\begin{equation}
\langle k,\bar q,\nu |\hat H_{T}^{nc-nc}|\varphi\rangle 
=e^{i\nu\varphi}t_{kq}\left( u_{k}v_{q}+
e^{2i\varphi}v_{k}u_{q}\right).
\label{eq:ce12}
\end{equation}
The phase dependent part of the associated second order
energy shift is thus
\begin{equation}
\Delta E_{nc-nc}^{(2)}(\varphi)=-2
\sum_{k,q}\, |t_{kq}|^2\,\frac{u_{k}v_{k}\, u_{q}v_{q}}
{E_{k}+E_{q}}\cdot\cos{2\varphi}\, ,
\label{eq:ce13}
\end{equation}
in perfect analogy to the result obtained for a superconducting
Josephson contact~\cite{amb63,ferrell88}, where
this is the only contribution.  Using the relation between the
average of $|t_{kq}|^2$ at the Fermi energy
and the normal state conductance, (21) directly leads to the 
result (8) at zero temperature.
For weakly interacting Bose gases, in turn, with the
explicit results (53) and (54) for the matrix elements given below,
the contribution (21) is easily shown to be 
smaller than the c-nc contribution (18) by a factor
$\sqrt{na^3}$~\cite{mei99}, and thus is negligible for 
gaseous BECs. 

For a given static value $\varphi$ of the phase difference between 
both condensates, the nondissipative current obtained from (6)
up to second order in $\hat H_{T}$ is thus of the form
\begin{equation}
I(\varphi)=-I_{c}\sin{\varphi}-J_{1}(0)\sin{2\varphi}\, ,
\label{eq:ce14}
\end{equation}
with a positive critical current $I_{c}=E_{J}/\hbar$. The
magnitude $J_{1}(0)<0$ of the second order contribution is
$-2/\hbar$ times the factor in parentheses in (18). 
The small second order term changes the  value of 
$\varphi$ at which the current is maximal from $\pi/2$
to $\pi/2+2|J_{1}(0)|/I_{c}$, however the critical current 
itself is unchanged to this order. In an
open system, with a reservoir of particles, where it is
possible to impose a constant external current, the phase
difference adjusts itself to a constant value, determined
by (22). This is the well known
dc-Josephson effect, i.e. a finite current at
vanishing chemical potential difference $\Delta\mu =0$.
In the case of two coupled BECs with a fixed total number
of particles, any current flow is connected with a 
finite value of $\Delta\mu$ even in the absence of an
external potential drop, because $\mu$ depends on the
particle number. Thus, by the Josephson relation (23),
one has inevitably a phase
difference which evolves in time. In addition,
finite dissipative currents appear, requiring a 
fully dynamical treatment, as will be given in the following 
sections.

\section{Josephson and normal currents at given $\Delta\mu$}

In the following, we want to determine the complete current in a 
situation with a finite difference in the chemical potentials. 
We start by considering the idealized case in which $\Delta\mu$ 
is considered as a fixed, externally given value. This approximation,
which applies to open systems like standard
Josephson contacts between superconductors, neglects the change in the
chemical potential associated with the transfer of particles 
in a system with a fixed total number of particles, an effect which
is taken up in section 5.  A finite value of $\Delta\mu$ gives
rise to an additional term $\Delta\mu\,\hat\nu$
to the total Hamiltonian (1). In close 
analogy to the calculation of currents in normal and 
superconducting tunnel junctions, this term may formally be
eliminated by a time dependent gauge transfomation. One thus obtains
a tunneling Hamiltonian with $\Delta\mu =0$,
in which the transfer matrix elements
are modulated in time with a factor $e^{i\varphi(t)}$ such that
\begin{equation}
\hbar \frac{d}{dt} \varphi(t) = - \Delta \mu .
\label{eq:cur1}
\end{equation}
Since the current operator 
\begin{equation}
\hat I=-\frac{d}{dt}\hat N_{a}=i\left( \hat{\Lambda}-\hat{\Lambda}
^{\dagger}\right)/\hbar
\label{eq:cur2}
\end{equation}
is already linear in the tunneling matrix elements, the problem
can, up to second order in $\hat H_{T}$, be formally treated like
in time dependent linear response~\cite{duk69}. 
It is then straightforward to show that the expectation value 
$I$ of the current is given by
\begin{eqnarray}
&& I = \frac{i}{\hbar} \langle e^{- i \varphi(t)} \hat{\Lambda} (t) 
- e^{i \varphi(t)} \hat{\Lambda}^{\dagger}(t) \rangle \nonumber \\
&& \hspace{0.2cm} + \frac{2}{\hbar^2} \Re \int_{-\infty}^t d 
t^{\prime} \, 
\bigl\{ e^{-i(\varphi(t)-\varphi(t^\prime))} \langle [\hat{\Lambda} 
(t), 
\hat{\Lambda}^\dagger (t^\prime)] \rangle  \label{eq:cur3} \\ 
&& \hspace{1.8cm}
+ e^{-i(\varphi(t)+\varphi(t^\prime))} \langle [\hat{\Lambda} (t), 
\hat{\Lambda} 
(t^\prime)] \rangle
\bigr\}\, , \nonumber
\end{eqnarray}
where the time dependence of the operators $\hat{\Lambda}$ and 
$\hat{\Lambda}^\dagger$ has to be taken with respect to the 
unperturbed Hamiltonian $\hat{H}_a + \hat{H}_b$. The expectation  
values reduce to ones in the associated ground state, provided
we restrict ourselves to the limit of zero temperature, as is
done throughout in the following.

As was pointed out above, the eigenstates of $\hat{H}_a + \hat{H}_b$
for gaseous BECs can be choosen as the Gross-Pitaevski wave
functions in each well, and a set of excited states obtained
from solving the Bogoliubov equations in local density
approximation. As long as the number of transferred
particles remains small compared to $\bar{N}_{a/b}$, these 
states are unaffected to lowest order. Thus both the 
condensate wave function as well as the
quasiparticle energies $E_{k}$ and amplitudes $u_k$ and $v_k$ 
are unchanged by the current flow.
The comparison with the numerical results in 
section 7 will show, that at least on the Gross-Pitaevski
level this approximation is well 
justified in the weak tunneling limit discussed here.

The first term on the right hand side of Eq.~(\ref{eq:cur3}) is then
\begin{eqnarray}
&& \frac{i}{\hbar} \langle e^{- i \varphi(t)} \hat{\Lambda} (t) - 
e^{ i 
\varphi(t)} \hat{\Lambda}^{\dagger}(t) \rangle \nonumber \\ && 
\hspace*{0.5cm} = 
- \frac{2 t_{cc}}{\hbar}
\sqrt{\bar{N}_a^c \bar{N}_b^c} \, \sin \varphi(t) = - I_c 
\sin \varphi (t)\, ,
\label{eq:cur4}
\end{eqnarray}
in perfect agreement with the result derived in the previous
section for a static situation.
The second term in Eq.~(\ref{eq:cur3})  
involves noncondensate tunneling. At fixed $\Delta \mu$, 
it is readily evaluated within Bogoliubov theory, giving three
contributions:
\begin{eqnarray}
&&  \frac{2}{\hbar^2} \Re \int_{-\infty}^t d t^{\prime} \bigl\{ e^{-
i(\varphi(t)-\varphi(t^\prime))} \langle [\hat{\Lambda} (t), 
\hat{\Lambda}^\dagger (t^\prime)] \rangle  \nonumber \\
&& \hspace{2.5cm}+ e^{-i(\varphi(t)+\varphi(t^\prime))} \langle 
[\hat{\Lambda} 
(t), \hat{\Lambda} (t^\prime)] \rangle
\bigr\} \nonumber \\
&& = J_n(\Delta \mu) - J_1 (\Delta \mu) \sin 2 \varphi(t) + J_2 
(\Delta \mu) 
\cos 2 \varphi (t).
\label{eq:cur5}
\end{eqnarray}
The first one is a phase independent normal current,  
accounting for the expected dissipative current flow 
towards the lower chemical potential. The $\sin{2\varphi}$-term 
describes a
second order Josephson current, associated with the
corresponding phase dependent coupling energy derived in the
previous section. Finally there is an interference current
proportional to $\cos{2\varphi}$ which is out of phase by
$\pi/2$ from the corresponding Josephson contribution. Similar
to the related $\cos{\varphi}$-term in superconductors~\cite{jos69}, 
this is a dissipative current which vanishes
at $\Delta\mu=0$, as the normal current does. Within the
local density approximation for solving the Bogoliubov-equations,
the current amplitudes $J_n$, $J_1$ and $J_2$ can be expressed
in terms of the tunneling matrix elements and 
the normal and anomalous spectral 
functions $A(k,\omega)$ and $B(k,\omega)$ of the 
weakly interacting homogeneous Bose gas.  In 
standard notation~\cite{fet71} they read
\begin{equation}
A(k, \omega)  = 2 \pi \bigl\{ u_k^2 \delta (\omega - 
E_k/\hbar) - v_k^2 
\delta (\omega + E_k/\hbar) \bigr\} 
\label{eq:cur6} 
\end{equation}
\begin{equation}
B(k, \omega)  = 2 \pi u_k v_k \bigl\{ \delta (\omega + 
E_k/\hbar) - 
\delta (\omega - E_k/\hbar) \bigr\}\, . 
\label{eq:cur7}
\end{equation}

With  $n_{B}(x)=(\exp{\beta x}-1)^{-1}$ as the Bose distribution 
function,
the current amplitudes in Eq.~(\ref{eq:cur5}) are then given by
\begin{eqnarray}
&& J_n (\Delta \mu) =  \bar{N}_{a}^c \sum_{q} \, 
\frac{|t_{cq}|^2}{\hbar^2} A (q, \Delta \mu/\hbar) \nonumber \\ 
&& 
\hspace{1cm} - \bar{N}_{b}^c \sum_{k} \, 
\frac{|t_{kc}|^2}{\hbar^2} A (k, - \Delta \mu/\hbar) \nonumber 
\\ && + 
 \sum_{k,q} \, \frac{|t_{kq}|^2}{\hbar^2} 
\int_{-
\infty}^{\infty} \frac{d \omega}{2 \pi} [n_B (\hbar \omega) - 
n_B(\hbar \omega + 
\Delta \mu)] \nonumber \\ && \hspace*{2cm} A(k,\omega) 
A(q,\omega + 
\Delta \mu/\hbar) \label{eq:cur8} 
\end{eqnarray}
and
\newpage
\begin{eqnarray}
&& J_2 (\Delta \mu) = \bar{N}_{a}^c \sum_{q} \, 
\frac{|t_{cq}|^2}{\hbar^2} B (q, -\Delta \mu/\hbar) \nonumber 
\\ && 
\hspace{1cm}  + \bar{N}_{b}^c \sum_{k} \, 
\frac{|t_{kc}|^2}{\hbar^2} B (k, - \Delta \mu/\hbar) \nonumber 
\\ && - 
 \sum_{k,q} \, \frac{|t_{kq}|^2}{\hbar^2} 
\int_{-
\infty}^{\infty} \frac{d \omega}{2 \pi} [n_B (\hbar \omega) - 
n_B(\hbar \omega + 
\Delta \mu)] \nonumber \\ && \hspace*{2cm}  
B(k,\omega) B(q,\omega + \Delta \mu/\hbar).
\label{eq:cur9}
\end{eqnarray}
The amplitude $J_1(\Delta \mu)$ of the second order Josephson
current can be obtained from that of the associated dissipative
contribution $J_2(\Delta\mu)$ via a Kramers-Kronig 
relation
\begin{equation}
J_1(\Delta \mu) = - {\cal P} \int_{-\infty}^\infty
 \frac{d \omega}{\pi} \frac{J_2 (\hbar \omega)}{\omega - \Delta 
\mu/\hbar}\, .
\label{eq:cur10}
\end{equation}

It is straightforward to check, that at zero temperature and 
vanishing $\Delta\mu$, the combination $\hbar J_{1}(0)/2$ is 
identical with the sum of the coefficients of
$\cos{2\varphi}$ in (18) and (21) of the second order
phase dependent coupling energy, as it should. 
As noted in the previous section, a particular feature of 
tunneling between Bose condensates is the appearance of
c-nc contributions. Indeed, as the first two terms in
Eq.~(\ref{eq:cur8}) show, they also contribute to the 
phase independent normal current, where they are in fact
dominant compared to the conventional nc-nc contribution
described by the last term. Another point that should be
mentioned, is the absence of a second order
contribution due to c-c  
tunneling~\cite{mei99}. This follows because the 
commutators involving only condensate operators  
in Eq.~(\ref{eq:cur3}) vanish.   
A different proof of this result may be 
obtained, by 
considering all possible Feynman diagrams associated with  
second order tunneling processes. Diagrams proportional to 
$|t_{cc}|^2$ are reducible and thus give no separate contribution.
  
The similarity of the second order currents in BECs and 
superconductors is a result of the close formal analogy of Bogoliubov 
and BCS theory. The additional contributions due to c-nc
tunneling in BECs lead to interesting new features, 
however. As will be 
discussed in section 6, for the Bose case the dissipative
currents $J_n (\Delta \mu)$ and 
$J_2 (\Delta \mu)$ remain finite at zero temperature 
in contrast to superconductors, where they
vanish exponentially due to the presence of an 
energy gap. These currents thus provide a mechanism for
the damping of Josephson oscillations even at $T=0$.
Similar to the situation in superconductors~\cite{zwe83},
it may be shown that the total dissipation associated with
the two irreversible currents is always positive~\cite{mei99}.
Physically, this property is equivalent to the statement, that
the sum $J_n + J_2 \cos 2 \varphi$ is strictly positive 
for $\Delta\mu >0$, i.e. the total dissipative current always flows  
towards lower chemical potential. It therefore leads
to a reduction of $\Delta \mu$ and hence to a damping of the phase 
dynamics, similar to the contribution of excited states in
driven two component condensates discussed in~\cite{villain99}.

\section{Path Integral description}

So far, we assumed $\hbar \dot{\varphi}$ to be given externally, thus 
neglecting 
the variation of the chemical potential with the total number of 
particles.  As we will see, however, this effect is quantitatively
important even in the regime where
$N_a-N_b \ll \bar{N}_{a/b}$. To lowest order in $\nu/\bar{N}_{a/b}$,
the dependence of the chemical potential on the actual number of 
particles can be accounted for by an 
additional `charging' energy $U 
(\hat{N}_a-\hat{N}_b)^2/8$, with 
$U=\partial \mu_a/\partial N_a+\partial \mu_b/\partial N_b$ being of 
order 
$\mu/\bar N$. In the limit $U \gtrsim \hbar I_c$, this term suppresses 
phase coherence, since by canonical quantization $\hat\nu 
\rightarrow i \partial/\partial\varphi\,$, charging 
effects 
are equivalent to quantum fluctuations of $\varphi$. For any $U \neq 
0$, the 
chemical potential difference in $\hbar \dot{\varphi} = -\Delta \mu$ 
contains an 
internal contribution $U\nu$ proportional to the number of 
transferred bosons. The resulting dynamics of $\varphi$ is therefore 
intrinsically nonlinear. Within a simple description,
these effects have already been introduced by Smerzi
et.al.~\cite{sme97},
who also took into account the $\nu$-dependence of the 
Josephson coupling $E_{J}\propto\sqrt{(\bar N_{a}+\nu)(\bar N_{b}
-\nu)}$. While this effect is indeed relevant for 
strong asymmetries $\nu = O(\bar{N}_{a/b})$, it is negligible
compared to the charging term of order $\mu\nu^2/\bar N$ provided
we are in the regime $E_{J}\ll \mu\bar N$ where Josephson
tunneling is a small perturbation.
In order to 
include charging effects in a microscopic description, we use the 
formalism developed by 
Ambegaokar, Eckern and Sch\"on \cite{amb82} for superconductor tunnel 
junctions.

In a coherent state path integral formulation for interacting bosons, 
the grand-
canonical partition function of two coupled condensates is given by
\begin{eqnarray}
&& Z = \int D \psi_{a}  D \psi_{b}   
\exp - \int_{0}^{\beta \hbar}\frac{ d \tau}{\hbar} \left[ 
\int d^3 x \, \psi_{a}^{\ast} (\hbar \partial_{\tau}-\mu_a) \psi_a  
\right. 
\nonumber \\
&& \hskip1cm \left. + \int d^3 y \, \psi_{b}^{\ast}(\hbar 
\partial_{\tau} -
\mu_b)\psi_{b} + H(\psi_a,\psi_b) 
 \right].  \label{grpart}
\end{eqnarray}
Here $\mu_{a/b}$ fix the average number $\bar N_{a/b}$ of
atoms in each condensate, while $H(\psi_{a},\psi_{b})$ is
the standard representation of the basic Hamiltonian (1),
in which the bosonic field operators are replaced by
complex c-number fields $\psi_{a/b}(x/y,\tau)$, depending
both on the spatial coordinates plus an imaginary time like
variable $\tau$~\cite{neg88}. 
With the charging term explicitly included in $H(\psi_a,\psi_b)$, 
Eq.~(\ref{grpart}) is also appropriate for describing a system with 
$N_a + N_b$ fixed. As in the analogous fermionic problem of
coupled superconductors, it is convenient
to introduce an auxiliary path integral $\int D V$ to remove the 
charging term 
quartic in the fields via a Hubbard-Stratonovich transformation. The 
functional 
integral is then evaluated in a saddle point approximation, 
the saddle point being just the condensate wavefunction for 
two 
separate condensates. Performing a gauge transformation, we may 
replace the 
integration variable $V(\tau)$ by the phase $\varphi(\tau)$, 
which is the dynamical variable of 
interest. Furthermore, we introduce Nambu spinors 
$\tilde{\Psi}^{\dagger}=(\tilde{\psi}_a^{\ast},\tilde{\psi}_a,\tilde{\psi}_b^{\ast},\tilde{\psi}_b)$ 
for the Gaussian fluctuations around the saddle point 
$\Phi^{\dagger}=(\phi_a^{\ast},\phi_a,\phi_b^{\ast},\phi_b)$. The
full partition function is then given by
\begin{eqnarray}
&& Z=  e^{-S_0} \int D\varphi \int D\tilde{\Psi} D 
\tilde{\Psi}^\dagger 
\nonumber \\
&& \hskip1cm
 \exp - \int_{0}^{\beta \hbar} \frac{d \tau}{\hbar} \,
\left[   \frac{\hbar^2 \dot{\varphi}^2}{2 U} - E_{J} \cos \varphi 
\right. 
\nonumber \\
&& \hskip0.5cm \left. + \frac{1}{2} \int d^3 x \, d^3 y \, 
\tilde{\Psi}^\dagger 
(-\underline{G}^{-1} + \tilde{t}) \tilde{\Psi} + \tilde{\Psi}^\dagger 
\tilde{t} 
\Phi 
+ \Phi^\dagger \tilde{t} \tilde{\Psi}
\right], \nonumber \\
&&  \label{grpart2}
\end{eqnarray}
where $S_0$ is a bulk contribution to the action. In coordinate
space, the $4 \times 4$ tunneling matrix 
$\tilde{t}$ is defined by
\begin{equation}
\tilde{t} = \left( \begin{array}{cc}
0 & \hat{t}_{xy} \\
\hat{t}_{xy}^{\dagger} & 0 
\end{array} \right), \hskip0.3cm 
\hat{t}   = \left( \begin{array}{cc}
t_{xy} & 0 \\
0 & t_{xy}^{\ast}  
\end{array} \right), \label{tmatrix}
\end{equation}
while the differential operator $\underline{G}^{-1}$ satisfies
\begin{equation}
\underline{G}^{-1} \left( \begin{array}{cc}
\underline{G}_{a} & 0 \\ 0 & \underline{G}_{b}  
\end{array} \right) = 
\left( \begin{array}{cc} 
\delta(x-x^\prime) \underline{1} & 0 \\
0 & \delta(y-y^\prime) \underline{1} 
\end{array} \right) \delta(\tau-\tau^\prime). \label{difop}
\end{equation}
In the last equation, $\underline{G}_{a/b}$ is shorthand for the 
noncondensate 
part of the $2 \times 2$ matrix Green's function for a 
weakly interacting Bose gas, as defined in~\cite{fet71}.

Completing the square in the exponent of (\ref{grpart2}), the 
Gaussian integral 
over $\tilde{\Psi}$ is readily performed and yields
\begin{equation}
Z = Z_a Z_b \int D \varphi(\tau) \exp (- S[\varphi]/\hbar). 
\label{grpart3}
\end{equation}
Here, $Z_{a/b}$ are the partition functions of the
separate condensates, while 
tunneling effects are taken into account by an effective action for 
the relative phase
\begin{eqnarray}
& S[\varphi] & = \int_{0}^{\beta \hbar} d \tau \, \left( 
\frac{\hbar^2 
\dot{\varphi}^2}{2 U} - E_{J} \cos \varphi \right) \nonumber \\
&& - \hbar \int_{0}^{\beta \hbar}  d\tau \, d\tau^\prime \left[
\alpha(\tau-\tau^\prime) \cos(\varphi-\varphi^{\prime})  \right. 
\nonumber \\ 
&& \hskip1.5cm \left.  + 
\beta(\tau-\tau^\prime) \cos(\varphi+\varphi^{\prime})\right], 
\label{effaction}
\end{eqnarray}
where $\varphi=\varphi(\tau)$ and $\varphi^\prime = 
\varphi(\tau^\prime)$. The 
integral kernels $\alpha(\tau)$ and $\beta(\tau)$ may be expressed in 
terms of 
the currents $J_n(\Delta \mu)$ and $J_2(\Delta \mu)$,
introduced in the previous section via their Fourier coefficients
\begin{eqnarray}
\alpha(i \omega_n)  & = & \int_{-\infty}^{\infty} \frac{d \omega}{2 
\pi} 
\frac{J_n(- \hbar \omega)}{i \omega_n - \omega} \label{eq:akernel} \\
\beta(i \omega_n)  & = & \int_{-\infty}^{\infty} \frac{d \omega}{2 
\pi} 
\frac{J_2(\hbar \omega)}{i \omega_n - \omega}\, , \label{eq:bkernel}
\end{eqnarray}
in an expansion of the $\beta\hbar$-periodic functions
\begin{equation}
\alpha(\tau)=\frac{1}{\beta\hbar}\sum_{n}\, e^{-i\omega_{n}\tau}
\alpha(i\omega_{n})
\end{equation}
in bosonic Matsubara frequencies $\omega_{n}=2\pi n/\beta\hbar$, $n$
integer.

The effective 
action Eq.~(\ref{effaction}) is formally identical with the one
obtained for a superconducting Josephson contact~\cite{amb82}, except
for the additional local contribution  
$-E_{J} 
\cos \varphi (\tau)$ describing the first order Josephson effect due 
to c-c tunneling. The terms nonlocal in time 
give rise to both the second order Josephson currents related to
$\beta (\tau - \tau^\prime)$ and the normal current associated
with $\alpha(\tau -\tau^\prime)$. 
For the idealized model discussed in section 6, where 
$J_{2}(\Delta\mu)$ and $J_{n}(\Delta\mu)$ are linear for 
small values of the chemical potential difference,
both $\alpha(\tau)$ and $\beta(\tau)$ asymptotically decay like 
$1/\tau^2$, i.e., the Josephson contact exhibits ohmic dissipation.

An exact evaluation of the path integral in (37) is 
obviously impossible, since the effective action is nonlocal
and - in particular - nongaussian. Nevertheless, as shown by
Ambegaokar, Eckern and Sch\"on~\cite{amb82}, it is a useful
starting point for the derivation of a semiclassical 
approximation for the phase dynamics. Indeed,
from the imaginary time description, it is possible to derive 
a semiclassical equation of motion in real time, 
which has the form of a classical 
Langevin equation with state dependent quantum noise.  
In the limit where $J_{2}(\Delta\mu)$ and the
normal current $J_{n}(\Delta\mu)=G_{n}\Delta\mu$ are purely ohmic,  
it reads
\begin{equation}
\frac{\hbar \ddot{\varphi}}{U} + G(\varphi)\hbar \dot{\varphi} + I_c \sin 
\varphi + J_1 
\sin 2 \varphi  = \eta_1 \cos \varphi + \eta_2 \sin \varphi,  
\label{qlangevin}
\end{equation} 
where $\eta_{1,2}$ are independent Gaussian random forces. Their 
autocorrelation 
function may be expressed in terms of the real time functions 
$\alpha(t)$ and $\beta(t)$ via
\begin{equation}
\langle \eta_{1/2} (t) \eta_{1/2} (t^\prime) \rangle = 2 \bigl[ 
\alpha^I(t-
t^\prime) \mp \beta^I (t-t^\prime) \bigr],
\label{eq:eta12}
\end{equation}
where $\alpha^I$ and $\beta^I$ are the imaginary parts of the analytic 
continuation of $\alpha(\tau)$ and $\beta(\tau)$ similar 
to~\cite{amb82}. The damping term is proportional to a 
phase dependent effective 
conductance $G(\varphi) = G_n(1 + \cos 2 \varphi)$, with a 
normal 
conductance $G_n$, which will be explicitly evaluated in section 6.  
It gives rise to a relaxational dynamics with  a typical 
time scale $1/G_n U$. Even without damping, however, the 
$\ddot{\varphi}$-term leads 
to currents which are not perfectly sinusoidal for a given external 
chemical potential difference. Neglecting the fluctuating forces,
the resulting average dynamics will be discussed in section 7 for
realistic values of the parameters in weakly coupled BECs.

\section{An explicit model}

If the trap potential close to the barrier varies slowly on the scale 
of the coherence length $\xi$, we 
may apply the approximations discussed in section 3, 
using eigenstates of a locally 
homogeneous system with constant external potential throughout
$a$ and $b$, respectively.
A simple model geometry for such a Josephson junction is shown in 
Fig.~\ref{fig:f1}:

\begin{figure}[t]
\epsfig{file=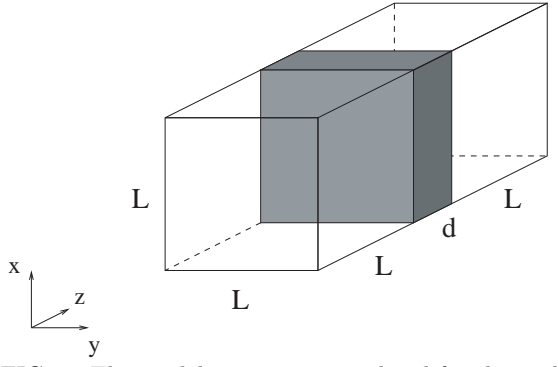,scale=0.40}
\caption{The model geometry considered for the evaluation of the 
current densities.} \label{fig:f1}
\end{figure}

Two BECs confined in a cubic volume $L^3$ are
separated by a square potential barrier of height $V_B$ and width 
$d$. For 
simplicity, we impose periodic boundary conditions in the directions 
parallel to the potential barrier. This simple model accounts
for the fact, that all currents must scale linearly with area.
It allows us to calculate the relevant currents per area
in terms of microscopic parameters of the bulk condensates. 
For a concrete experimental setup, one may then determine
the actual critical current $I_{c}=j_{c}A$ from the
associated current densities and the effective contact area $A$.

For this system, the currents $I_c$, $J_n$, $J_1$ and $J_2$ 
may be evaluated explicitly from the detailed form of the 
tunneling matrix elements. Since we have assumed translation
invariance parallel to the barrier, the associated momenta
$\vec k_{\parallel}\, ,\vec q_{\parallel}$ are conserved, 
and the problem is effectively one-dimensional (1d). The
tunneling amplitudes are then obtained from the current
matrix elements~\cite{duk69}
\begin{equation}
t_{lr}=\frac{\hbar^2}{2m}\left[ \chi_{l}\frac{d\chi_{r}}{dz}-
\chi_{r}\frac{d\chi_{l}}{dz}\right]_{z=0}\cdot
\delta_{\vec k_{\parallel},\vec q_{\parallel}}
\label{eq:ev1}
\end{equation}
of the 1d wave functions $\chi_{l,r}(z)$, taken at the 
center of the barrier at $z=0$. In the limit of weak
tunneling, the barrier height $V_{B}$ is much larger than
the average chemical potential $\bar\mu=(\mu_{a}+\mu_{b})/2$.
The wave functions in (44) are therefore effectively
{\it single particle} eigenstates which decay exponentially
like $\exp{\pm\kappa_{\mu}z}$, with inverse characteristic
length $\kappa_{\mu} = \sqrt{2m (V_B - \mu)/\hbar^2}$.
For a small difference $\Delta\mu=\mu_{a}-\mu_{b}\ll\bar\mu$
in the chemical potentials, (44) then reduces to
\begin{equation}
t_{lr}=\frac{\hbar^2\kappa_{\bar\mu}}{m}\,\chi_{l}(0)\chi_{r}(0)
\cdot\delta_{\vec k_{\parallel},\vec q_{\parallel}}.
\label{eq:ev2}
\end{equation}
The calculation of the different tunneling amplitudes
$t_{cc}$, $t_{kc}$ and $t_{kq}$, thus requires
\begin{itemize}
	
\item a solution of the 1d GPE for a finite barrier, which 
approaches $\phi_{c}=\sqrt{N_{c}}\cdot\chi_{c}\to\sqrt{n_{c}}$
far from the barrier, i.e. a uniform condensate with density
$n_{c}$ (note that the boundaries at $z=\pm(L+d/2)$ are
irrelevant since $L\gg\xi$), and

\item corresponding solutions of the 1d Bogoliubov equations
with finite momenta $k\, ,q$, which smoothly
connect to the exponentially decaying single particle states
below the barrier.

\end{itemize}
For concreteness we consider the eigenstates $\chi_{l}$ in
condensate $a$, to the left of the barrier. It is then 
straightforward to show that
\begin{eqnarray}
\chi_{c,a} (z) & = & \frac{1}{\sqrt{L}} \frac{\kappa_{\mu_a} 
\xi_a}{\sqrt{2}}
\bigl( \sqrt{1 + \frac{2}{(\kappa_{\mu_a} \xi_a)^2}} - 1\bigr) e^{-
\kappa_{\mu_a} (z + d/2)} \nonumber \\
 & & \stackrel{V_B \gg \mu_a}{\rightarrow}  \frac{1}{\sqrt{L}} 
\frac{1}{\sqrt{2} \kappa_{\mu_a} \xi_a}  e^{-\kappa_{\mu_a} (z + 
d/2)}
\label{eq:ev3}
\end{eqnarray}
provides an appropriate solution of the GPE.
It is obtained by connecting the well known
solution $-\sqrt{n_{c}}\tanh{\frac{z+\delta}{\sqrt{2}\xi}}$
of the full GPE in the regime $z<-d/2$ to 
an exponentially decaying solution of the {\it linearized}
equation below the barrier $z>-d/2$. The condition
that $\chi$ and its first derivative are continuous at $z
=-d/2$ fixes the prefactor in the solution of the linear
equation. The linearization is justified in the limit
$\kappa\xi\gg 1$, where the mean field interaction is negligible
compared to the repulsive potential separating the two
condensates. Using the solution (46) and the condition
$\Delta\mu\ll\bar\mu$, the c-c tunneling amplitude is given by
\begin{equation}
t_{cc}=f(V_{B}/\bar\mu)\cdot
\frac{\sqrt{\mu_{a}\mu_{b}}}{\kappa_{\bar\mu}L}\,
e^{-\kappa_{\bar\mu}d}\, ,
\label{eq:ev4}
\end{equation}
with a correction factor
\begin{equation}
f(x)= \left( 1-(x-\sqrt{x^2-1})\right)^2\, ,
\label{eq:ev5}
\end{equation} 
which is smaller than one and approaches unity in the high barrier
limit $V_{B}\gg\bar\mu$. Since $t_{cc}$ is proportional to $1/L$,
the resulting Josephson coupling 
energy $E_{J}=2t_{cc}\sqrt{\bar N_{a}\bar N_{b}}$ scales like the
contact area $A=L^2$, as it should. The associated critical current
density is
\begin{equation}
j_{c}=2f(V_{B}/\bar\mu)\,\frac{(\mu_a n_{c,a}\,\mu_{b} n_{c,b})^{1/2}}
{\hbar\kappa_{\bar{\mu}}}\cdot e^{-\kappa_{\bar\mu}d}. 
\label{eq:ev6}
\end{equation}
As expected, it is linear in the barrier transmission amplitude
$\exp{-\kappa_{\bar\mu}d}$. More interesting is the prefactor which,
by the standard relation $2p=\mu n_{c}$ for the ground state
pressure of a weakly interacting Bose gas, 
is just proportional to the geometric average of 
the ground state pressures $p_a$ and $p_b$ of the two condensates. 
It is instructive to compare this with the corresponding
result for coupled {\it ideal} Bose gases, where the 
condensate wave functions are the well known wavefunctions of the one 
particle 
ground state in a potential well of finite height. 
The associated c-c tunneling amplitude is then readily evaluated,
giving
\begin{equation}
t_{cc}^{(0)} =  2 \pi^2 \frac{\hbar^2}{m} \frac{1}{\kappa_{\bar{\mu}} 
L^3} e^{-
\kappa_{\bar{\mu}} d}.
\label{eq:ev7}
\end{equation}
>From this, we immediately find a critical current
\begin{equation}
I_c^{(0)} = 4 \pi^2 \frac{\hbar \sqrt{\bar{n}_a^c \bar{n}_b^c}}{m 
\kappa_{\bar{\mu}}} 
e^{-\kappa_{\bar{\mu}} d}
\label{eq:ev8}
\end{equation} 
which is independent of the contact area.  
Evidently, the behaviour implied by Eq.~(\ref{eq:ev8})  
is an artefact of the ideal Bose gas, which  
exhibits a vanishing quantum pressure below the 
condensation temperature $T_c$ in the thermodynamic limit.  The 
critical current density thus remains finite for large contact areas
only if the repulsive interaction between the particles is included.
The linear scaling of the
critical current with area is also found in the model 
by Zapata {\it et al.}~\cite{zap98}. In fact, our 
result for the critical current agrees with theirs for
a plane barrier to leading order 
in $e^{-\kappa_{\bar{\mu}}d}$, i.e. with $f=1$. As a final point of
our discussion of c-c tunneling, we mention that due to the
$N$-dependence of the chemical potential in (47), the amplitude
$t_{cc}$ - which is identical with the 
coupling parameter $K$ introduced in~\cite{sme97} - depends on
the number of particles in the condensate. In the limit
$|\nu|\ll\bar N_{a/b}$ discussed here, this effect is 
negligible compared to the charging effects, as mentioned above. 
In situations, however, where $|\nu|$ becomes itself of order
$N_{a/b}$, this dependence is important and the c-c tunneling
amplitude cannot be treated as a constant, as assumed in 
ref.~\cite{sme97}.

To calculate the remaining amplitudes $t_{kc}$ and $t_{kq}$
which enter into the higher order and dissipative currents,
we need to match a solution of the 1d Bogoliubov equation
for a homogeneous condensate in $z<-d/2$ to an exponentially
decaying wave function below the barrier. Since the condensate
density $n_{c}(z)$ there vanishes very quickly like $\exp{-2\kappa z}$,
the mixing between the amplitudes $u_{k}(z)$ and $v_{k}(z)$
may be neglected (for simplicity, $k$ and $q$ here are
just the $z$-components of the original 3d wave vectors $k$ and $q$).
Connecting the trivial solution $\left( 2/L\right)^{1/2}
\sin{(kz+\delta)}$ for $u_{k}(z)$ in $a$, to one below the barrier
such that the function and its first derivative are continuous
at $z=-d/2$, we find
\begin{equation}
\chi_{k}(z)=\left(\frac{2}{L}\right)^{1/2}\frac{k}{\kappa_{\mu_{a}}}
\cdot e^{-\kappa_{\mu_a} (z + d/2)}.
\label{eq:ev9}
\end{equation}
Here, it has been assumed that $k\ll\kappa$, i.e. 
the kinetic energy for motion 
perpendicular to the barrier is small compared to $V_{B}$.
The tunneling amplitudes involving noncondensate states now 
follow easily from (45) and are given by
\begin{equation}
t_{kc}=\frac{\hbar^2}{m\xi_{a}\kappa_{\bar\mu}L}\,
\frac{k}{\sqrt{1+(k\xi)^2}}e^{-\kappa_{\bar{\mu}} d} 
\cdot\delta_{\vec{k}_{\parallel},\vec{0} } 
\label{eq:ev10}
\end{equation}
and
\begin{equation}
t_{kq}=\frac{2\hbar^2}{m\kappa_{\bar\mu}L}\,
\frac{kq}{\sqrt{1+(k\xi)^2}\sqrt{1+(q\xi)^2}}e^{-\kappa_{\bar{\mu}} d} 
\cdot\delta_{\vec{k}_{\parallel},\vec{q}_{\parallel}}\, , 
\label{eq:ev11}
\end{equation}
again assuming $\Delta\mu\ll\bar{\mu}$. These matrix elements
vanish linearly with the incoming momentum at low energies,
as expected for a tunneling amplitude. In order to avoid
the related unlimited increase of the amplitudes for large
momenta, we have introduced a cutoff $1/\sqrt{1+(k\xi)^2}$
which leads to a saturation of the matrix elements at an
energy scale still much smaller than the barrier height $V_{B}$.
In fact for energies in this range, the transfer
Hamiltonian fails, because the coupling between the two sides
can no longer be treated perturbatively. Fortunately, for
small chemical potential differences $\Delta\mu\ll\bar\mu$,
only the low energy phonon like excitations with sound
velocity $c=\sqrt{\mu/m}$ contribute to $J_{n}$ and $J_{2}$.
Indeed, for small frequencies, the
spectral functions are antisymmetric in 
$\omega$ and are equal up to a sign
\begin{equation}
A(k,\omega)\simeq -B(k,\omega)=
\frac{\pi\mu_{a}}{\hbar\omega}\,\delta(\omega -c_{a}k)\quad
(\omega>0).
\label{eq.:ev12}
\end{equation}
The sum over momenta is thus restricted to a regime where the 
cutoff is irrelevant. It plays a role only for the less
important second order Josephson current $J_{1}$, which
is small compared to the leading first order term, with a 
typical magnitude
\begin{equation}
J_{1}(\Delta\mu =0)\approx I_{c}\cdot e^{-\kappa_{\bar\mu}d}/
{\kappa_{\bar\mu}\xi}.
\label{eq:ev13}
\end{equation}
Using (55), the c-nc contributions to 
$J_n$ and $J_{2}$ are readily evaluated in the limit $L 
\rightarrow \infty$, where the one particle energy spectrum 
becomes continuous. To lowest order in the chemical potential
difference, both currents are linear in $\Delta\mu$,
behaving like
\begin{equation}
J_n (\Delta \mu\to 0) =J_2 (\Delta \mu\to 0)=G_{n}\Delta\mu .
\label{eq:ev14}
\end{equation}
The associated normal conductance per area for a symmetric 
situation is finite at zero temperature and given by
\begin{equation}
G_{n}/A =	\frac{2\sqrt{2}n_{c}}{\hbar 
\kappa_{\bar{\mu}}^2\xi}\cdot e^{-2 \kappa_{\bar{\mu}} d}.
\label{eq:ev15}
\end{equation}
Concerning the nc-nc terms, it is straightforward to see that 
their contribution to $G_{n}$ vanishes like $T^4$ and thus is 
negligible. The same applies to the nc-nc contribution to
$J_{1}(\Delta\mu =0)$ which, as noted in section 3, turns out 
to be smaller than the c-nc term (56) by a factor 
$\sqrt{na^3}$~\cite{mei99}. 

A remarkable feature of tunneling between Bose condensates is 
the presence of a normal conductance proportional to the
condensate density. Contrary to the situation encountered in 
superconductors, the dissipation in a Bose Josephson 
junction thus remains finite 
even at $T=0$. It is obvious that this result relies on
the continous excitation spectrum of the homogeneous system,
and thus its relevance to BECs in traps might be questioned.
I practice, though, for $\Delta\mu\gtrsim\hbar\omega$ and for
the experimentally relevant temperatures where $k_{B}T\gg
\hbar\omega$, the discreteness of the spectrum in the
harmonic traps is irrelevant, and the continuum approximation
thus well justified. We shall see, in the next section, 
that dissipative currents play an important role in the
dynamics of coupled BECs, even if they are small.

\section{Dynamics for realistic parameters}

On the basis of our microscopic model developed in the previous
sections, we now aim to provide realistic estimates for the
necessary requirements and the expected dynamics in a Bose 
Josephson junction, which may be realized experimentally
with present techniques. Assuming a contact area  
$A=20 \,\mu$m$^2$ and a condensate extending about $10\,\mu$m
in the direction perpendicular to the barrier, we have 
roughly $\bar N_{a}=\bar N_{b}=10^5$ atoms on each side 
on average. With the known mass and scattering length the 
associated mean density $n=5 \times 10^{14}\,$cm$^{-3}$ then
determines the average chemical potential $\bar{\mu} = 
h \times 8.2$ kHz for $^{23}$Na, which is the more
favorable case for observing tunneling due its lighter
mass compared with $^{87}$Rb. Since the tunneling amplitude
decreases exponentially with barrier thickness $d$, it is
desirable to have widths $d$ as small as possible.  With 
barriers realized by a blue detuned laser, $d=1\,\mu$m
is difficult to reach, but still a realistic value. Using (49),
the resulting critical current can now be determined for any
given barrier height. With a large barrier $V_{B}=5\bar\mu$, 
however, the resulting critical current is only $I_{c}=584\,$sec$^{-1}$,
which is far too small to be observable. Indeed, the
corresponding tiny Josephson coupling energy is not sufficient to 
establish phase coherence across the barrier against even the 
minute thermal fluctuations at the typical temperatures of 
gaseous BECs. In fact, these fluctuations give rise to a  
smearing $\langle \Delta\varphi^2\rangle =k_{B}T/E_{J}$ of the
relative phase. A Josephson effect, which requires small
phase fluctuations
$\langle \Delta\varphi^2\rangle\ll 1$, is thus only possible if
$I_{c}\gg k_{B}T/\hbar\simeq 10^4\,$sec$^{-1}$ at $T=100\,$nK.
For conventional superconductors, where $T$ is of the order of $1\,$K, 
this condition expresses the well known fact that minimum critical 
currents are around $1\,\mu$A~\cite{tinkham96}.

In order to observe Josephson tunneling in gaseous BECs, one 
therefore needs barriers which are not much larger than the
average chemical potential. For a quantitative estimate, it 
is important to consider the necessary magnitude of the 
oscillating currents, which are detectable as a clear signal
of the Josephson effect against experimental noise.
For a given difference $\Delta \mu$ in the chemical
potentials, the amplitude of the oscillation in the number of
particles is $E_J/\Delta \mu$. With current experimental resolution,  
this amplitude should be at least around five percent of the total  
particle number to be detectable. Using typical
values $\Delta\mu=0.02\bar\mu$, this requires Josephson coupling 
energies $E_J\gtrsim 10^{-3}(N_a + N_b)\bar\mu$ (note that 
this is still consistent with the requirement $E_J \ll(N_a + 
N_b)\bar\mu$ of a tunneling Hamiltonian description). With
the numbers above, this translates into critical currents
$I_{c}\gtrsim 10^7\,$sec$^{-1}$, putting a much stronger limit
than that of negligible thermal fluctuations. To realize a
Josephson contact in the required range, it is necessary to choose a 
rather small barrier height $V_{B}=1.25 \bar{\mu}$. The 
critical current in a condensate with the above parameters 
is then around $4\cdot 10^6\,$sec$^{-1}$.
For $^{23}$Na, the observation of a Josephson effect is therefore 
close to the limit of detectability, and may be possible in an
optimized setup. By contrast, for $^{87}$Rb, the achievable 
critical currents even for barriers as low as $1.25\bar\mu$
are an order of magnitude smaller than those in $^{23}$Na.
Thus it appears unlikely, that coherent transfer of atoms 
across a tunnel barrer may be realized in this case with 
presently available condensates. 

For a quantitative analysis, it is necessary to consider the
time evolution of the relative particle number $\nu=(N_{a}
-N_{b})/2$ including `charging' effects and dissipation.
As noted in section 5, the total difference $\Delta\mu$
in the chemical potential is a sum of a possible external
contribution $\Delta\mu_{0}$ (induced e.g. by a difference
in the gravitational potentials as realized in~\cite{and98})
plus an internal `charging' term $U\nu$ associated with a
difference in the densities due to particle transfer.
The generally valid Josephson  relation (23) thus reads
\begin{equation}
\hbar\dot\varphi=-\Delta\mu_{0}-U\nu\, .
\label{eq:dyn1}
\end{equation}
For a constant $\Delta\mu_{0}$, a further time derivative
then leads to the semiclassical equation of motion
\begin{eqnarray}
&& \frac{\hbar}{U} \ddot{\varphi} (t) + G_n \hbar \dot{\varphi} (t) 
(1 + \cos 2 
\varphi (t)) \nonumber \\ && \hspace{0.5cm} + I_c \sin \varphi (t) + 
J_1 (- 
\hbar \dot{\varphi} (t)) \sin 2 \varphi (t) = 0 \label{eq:dyn2}
\end{eqnarray}
provided $-\dot\nu$ is replaced by the total current obtained for
a given instantaneous $\Delta\mu=-\hbar\dot\varphi$. Obviously
this is just Eq.(42) without the fluctuating forces, determining
the dynamcis of the {\it averaged} relative phase and the 
associated mean current. Eq.(60) is the analog of the well known
resistively shunted junction model for superconducting 
Josephson junctions~\cite{tinkham96}, and is valid for small 
phase fluctuations $\langle\Delta\varphi^2\rangle \ll 1$. Apart from the
condition $E_{J}\gg k_{B}T$ mentioned above, this also 
requires that the Josephson coupling is much larger than
the charging energy $U$. Now, in a symmetric situation with
$\bar N_{a}=\bar N_{b}=\bar N$ and for the parameters choosen
above, $U=2\bar\mu/\bar{N}$ is only of order
$h\times 0.16\,$Hz compared with $E_{J}=h\times 0.6\,$MHz.
Nevertheless, charging effects are far from negligible, since
for small values of $\Delta\mu_{0}$ they are the dominant
contribution in (59) driving the phase dynamics. To give a
quantitative example, we have numerically integrated the 
averaged, semiclassical time evolution of $\varphi$ and $\nu$ 
for the above parameters, neglecting second order currents.
As is shown in Fig. 2, the normalized particle number 
difference $z=\nu/\bar N$
for a constant offset potential $\Delta\mu_0 = 0.03\bar\mu$
and initial condition $\nu(0)=0$ exhibits appreciable deviations
from a pure sinusoidal oscillation predicted in the absence
of `charging' effects. Obviously both the
amplitude and the frequency of the Josephson oscillations
are modified, although only by a factor of order of unity. 
The typical time scale for the oscillations 
is msec, which is well within the reach of experimental 
observation.

\begin{figure}
\centering\epsfig{file=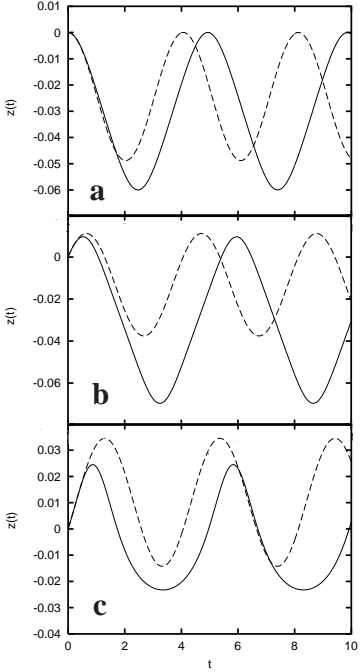,scale=0.65}
\caption{Nonlinearity of the Josephson oscillations due 
to charging 
effects. For the parameters discussed in the text, the 
normalized population difference is shown for initial 
relative phases 
$\varphi = 0,1,2$ (from top) both with (solid line)
or without charging effects (dashed line). 
The time is measured in msec.} 
\label{fig:f2}
\end{figure}

Regarding the higher order and dissipative currents, which 
have been neglected in Fig. 2, it turns out that for the low 
barriers choosen here, the amplitude of the second order 
Josephson current is not much smaller than the leading 
term, with a typical value
$J_1(\Delta \mu=0)\simeq 0.1I_{c}$. Now although this 
contribution oscillates at twice the frequency of the
first order term, an experimental detection is probably out of  
reach at present, given the narrow margin for seeing even the
dominant effect, as discussed above.
The dissipative normal and interference 
currents, however, significantly change the dynamics of the 
coupled condensates and thus should be indirectly observable 
even if the individual contributions are tiny. Indeed from  
Eq.~(\ref{eq:dyn2}) 
it is evident, that these currents lead to a damping of 
the phase dynamics with a 
typical time scale of order $\tau = 1/G_n U$. For the parameters 
choosen above, the normal conductance is  $G_{n}=250/h$,
giving a relaxation time $\tau = 0.025 s$. The amplitude of the Josephson 
oscillations will therefore decay to $1/e$ of their inititial value 
after typically 25 periods. As pointed out by Ruostekoski
and Walls÷\cite{walls98}, dissipative currents also  
eliminate the macroscopic quantum self trapping predicted 
by Smerzi et.al.~\cite{sme97,kleber71} for 
systems with a large initial population imbalance and small
Josephson coupling. For a quantitative estimate, we consider the influence
of an ohmic normal current $J_{n}=G_{n}\Delta\mu$ in a
simplified equation of motion for the normalized
particle number difference $z$. With $\tilde t=2t_{cc}t/\hbar$
as a dimensionless time and for a symmetric situation with
$\Delta\mu_{0}=0$ they read
\begin{equation}
\frac{dz}{d\tilde t}=-\sqrt{1-z^2}\sin{\varphi}-
2\hbar G_{n}\bar\mu/E_{J}\cdot z
\label{eq:dyn3}
\end{equation}
and
\begin{equation}
\frac{d\varphi}{d\tilde t}=-\Lambda z-
\frac{z}{\sqrt{1-z^2}}\cos{\varphi}\, .
\label{eq:dyn4}
\end{equation}
Here $\Lambda=U\bar N/2t_{cc}=\bar\mu/t_{cc}$ is  
essentially the ratio
between the condensate ground state energy $E_{0}\approx
\bar\mu\bar N$ and the Josephson coupling energy $E_{J}=
2t_{cc}\bar N$. The $\cos{\varphi}$-contribution in (62)
arises from the $z$-dependence of the critical current
$I_{c}(z)=I_{c}\cdot\sqrt{1-z^2}$, relevant for large
asymmetries $z=O(1)$, while the second order Josephson currents
discussed above are neglected for simplicity~\cite{note1}. The
phenomenological equations (61,62) lead to a self trapping of
the condensate, provided $\Lambda$ is larger than a critical
value which depends on the initial asymmetry $z(0)$. In the
absence of dissipation, the asymmetry will be maintained
in time, leading to a behaviour as shown in Fig. 3 (dotted line),
with $z(t)$ oscillating near its initial value $z(0)=0.6$ (as
in ref.~\cite{sme97} we choose $\Lambda=11$ and $\varphi(0)=0$).
\begin{figure}
\epsfig{file=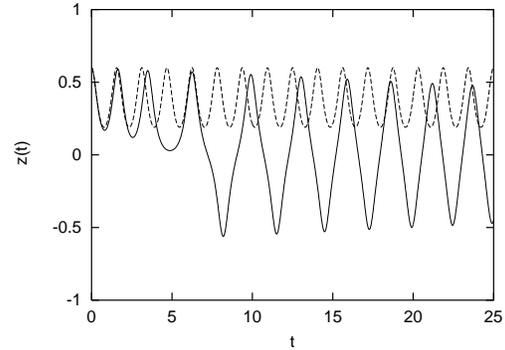,scale=0.45}
\caption{Destruction of macroscopic quantum 
self trapping through dissipative currents.
For comparison, the dynamics obtained without 
dissipation is also shown (dashed line).
Time is measured in units of $\hbar/2t_{cc}$ and thus with
a typical value $t_{cc}=h\times 20\,$Hz, $t=25$ corresponds
to about $0.1\,$sec}
\label{fig:f3}
\end{figure}
It is obvious that normal currents, which lead to an
equilibration of the chemical potentials $\Delta\mu=\mu_{a}-
\mu_{b}\to 0$ for long times, will eventually destroy a self trapped 
state. Observation of macroscopic quantum self trapping is therefore
possible only if the time scale $1/G_{n}U$ for equilibration is
much larger than $\hbar/2t_{cc}$, which is the typical scale for
the dynamics in the trapped state. This requires that the conductance 
obeys $\hbar G_{n}\ll E_{J}/2\bar\mu$, i.e.
the coefficient in front of the dissipative term
in (61) should be small compared to one. Now from our 
microscopic results (49) and (58), we find that
\begin{equation}
2\hbar G_{n}\bar\mu/E_{J}=\frac{2\sqrt{2}}{f(V_{B}/\bar\mu)
\kappa_{\bar\mu}\xi}\cdot e^{-\kappa_{\bar\mu}d}\, .
\label{eq:dyn5}
\end{equation}
For high barriers, with a corresponding tiny Josephson 
coupling energy, this ratio can certainly be made
small enough, such that dissipation has essentially no 
influence on the dynamics. 
As pointed out above, however, realistic systems allowing to
observe Josephson tunneling require {\it small}
barriers. For those, the inequality is strongly violated
and in fact, for the parameters choosen above, the ratio
in (63) is close to one.
For a quantitative example, which allows a comparison with
the results obtained in ref.~\cite{sme97}, we assume
a condensate with much lower density 
$n\approx 1.3\cdot 10^{13}\,$cm$^{-3}$ and a comparatively large
value $t_{cc}=h\times 20\,$Hz, giving $\Lambda=11$~\cite{note2}.
With reasonable parameters $V_{B}\approx 2\bar\mu$, $\kappa_{\bar\mu}d
=4.2$ and $\kappa_{\bar\mu}\xi=3$, the ratio
in (63) has a rather small value $0.0275$.
The resulting time evolution of $z(t)$
including dissipation, is shown as the solid line in Fig. 3.
Evidently self trapping is destroyed rather quickly even in this case
and in fact, the situation does not change qualitatively
for other parameters which seem accessible. From our 
microscopic results, it thus appears unlikely
that macroscopic quantum self trapping can actually 
be observed in condensates which are realizable at present.  

\begin{figure}
\epsfig{file=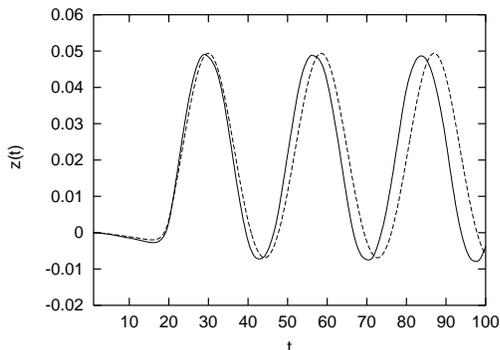,scale=0.45}
\caption{Comparison of the normalized particle number difference $z(t)$ 
obtained by a numerical 
integration of the GPE (solid line) and from the semiclassical 
dynamics based on the tunneling Hamiltonian
(dashed line) for a high and narrow potential barrier with $d=2 
\xi$ and $V_B 
= 3 \bar{\mu}$} \label{fig:f4}
\end{figure}

Finally we want to discuss the problem, to which extent
the tunneling Hamiltonian is still applicable in a regime,
where the barrier $V_{B}$ is only slightly larger than
the chemical potential, as was assumed above.  Now
at least on the mean-field-level, 
the quality of the transfer Hamiltonian model may be 
tested by comparing its predictions 
with those obtained from a numerical integration of 
the time dependent GPE in the geometry 
of Fig.~\ref{fig:f1}. Due to the periodic boundary conditions in 
the directions parallel to the potential barrier, the 
problem is effectively one-dimensional. 
The time evolution of the condensate wavefunction can thus be 
easily determined 
using a split-operator Fourier technique. In Fig.~\ref{fig:f4}, the 
normalized number of particles which have tunneled through the
barrier is shown for a case with $d= 2 \xi\approx 0.3\,\mu$m
and a comparatively large value $\Delta \mu_0 = 0.2 \bar{\mu}$.
Time is measured in units $\hbar/\bar{\mu}$,
corresponding to about $0.6\,$msec at $t=30$ for our values above.
The height
of the potential barrier is lowered continuously from an initial 
value $10\bar\mu$

\begin{figure}
\epsfig{file=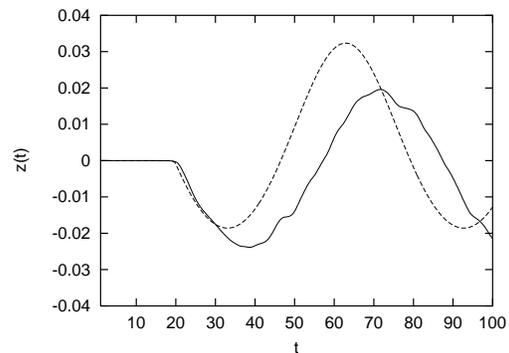,scale=0.45}
\caption{Same as Fig.4, but now for a barrier with $d=5 
\xi$ and $V_B 
= 1.4 \bar{\mu}$} \label{fig:f5}
\end{figure}

\hspace{-0.52cm}
at $t=10$ to $V_{B}=3\bar{\mu}$ at $t = 20$.
Charging effects are contained intrinsically in the GPE or via the
$U\nu$-contribution to the chemical potential difference; the
initial conditions are $\nu(0)=\varphi(0)=0$.
Evidently, the result obtained within the transfer Hamiltonian model 
is in excellent 
agreement with that from a numerical solution of the GPE, 
in the limit of a high and narrow barrier.
Note that there are no adjustable parameters here, 
with all quantities being determined by the 
microscopic parameters of the Bose system.
Remarkably, the transfer Hamiltonian still provides a 
reasonable approximation
even for the case of relatively wide and low 
potential barriers, which are  relevant experimentally. 
For example, in Fig.~\ref{fig:f5}, the function $z(t)$ is 
plotted for an identical situation as above, however with 
$d= 5 \xi\approx 0.8\,\mu$m, $V_B = 1.4 \bar{\mu}$ and 
$\Delta \mu_0 = 0.1 \bar{\mu}$
Although the 
semiclassical dynamics based on the transfer Hamiltonian does 
not reproduce the higher harmonics present in the solution of  
the GPE, which are essentially a result of the nonadiabatic 
lowering of the barrier, it still provides a reasonable
approximation even in this rather extreme case of a barrier
which is only slightly larger than $\bar\mu$.

\section{Conclusions and outlook}

In summary, we have developed a microscopic theory of Josephson 
tunneling in weakly interacting BECs. It is essentially the analog
within Bogoliubov theory of the standard work by Ambegaokar $et.al$.~\cite{amb63,amb82}
on the tunneling Hamiltonian description of
the Josephson effect between BCS superconductors. Apart from the
fact that the Bogoliubov theory only applies far below the
condensation temperature $T_{c}$, while BCS is valid right up
to $T_{c}$, there are a number of further crucial differences.
Most importantly, the Josephson effect in BECs arises already in
{\it first} order in the tunneling amplitude. As a result, the
Josephson current dominates any other contributions in the 
limit of weakly coupled condensates, to which our discussion
has been restricted. Nevertheless, the second order dissipative
currents have a strong influence on the dynamics of coupled
condensates, because they remain finite even at zero temperature
due to the absence of a gap. Remarkably, the dominant contribution
to the normal current, which leads to a damping of the Josephson
oscillations, arises from c-nc tunneling and is thus proportional
to the condensate density. The explicit calculation of the total
current has been performed here in a very simplified model. It has
the advantage, however, in providing analytical and physically
transparent results, allowing to determine 
the currents in a concrete experimental realization 
from the knowlegde of the bulk condensate properties,
the barrier height and the effective contact area. The fact that
tunneling in a Bose system is possible at all energies, made it 
necessary to specify the associated matrix elements in much more
detail than in superconductors. There, only the Fermi energy is
relevant, and thus the matrix elements can be replaced by a 
constant which is fixed by the normal state conductance. The
effect of `charging' for condensates with a constant total
particle number has been shown to be quantitatively important 
in realistic situations, even though the relevant energy $U$
is much smaller than the Josephson coupling energy $E_{J}$.

Regarding the prospects for an experimental observation of the Josephson
effect in condensates of dilute atomic gases, our estimates
show that this requires barriers which are only slightly higher
than the chemical potential. With presently available condensates,
the observation appears to be possible with $^{23}$Na, not, however,
with heavier atoms. One of the problems,
for instance, which may suppress the small oscillating
Josephson currents, are random fluctuations in the barrier height
due to fluctuations in the laser intensity, a complication 
which has not been considered so far.

On the theoretical side, quantitative calculations
for realistic geometries of coupled traps (in particular beyond
the Gross-Pitaevski level) would be useful and also a careful consideration of 
finite temperature effects. For example, the currents due
to particles in the thermal cloud have been estimated in~\cite{zap98}, 
however more work needs to be done in this
direction. Finally, since Josephson oscillation frequencies are 
typically of the same order than those of collective modes
\cite{str98} of the individual condensates, it is important 
to investigate a possible coupling between the intra- and
inter-well dynamics.

We gratefully acknowledge helpful discussions with T. Esslinger and I. 
Bloch on the experimental aspects of our work.


\begin{references}
\bibitem{note1} As mentioned in section 6, for $z=O(1)$ the 
tunneling amplitude $t_{cc}$ is itself $z$-dependent, an effect 
which is not taken into account here. Note also, that our $\varphi$
has the opposite sign as the one in ref.~\cite{sme97}.
\bibitem{note2} The value $\Lambda=11$ is rather small, 
for our parameters above $\Lambda$ is in fact around $2500$.
\end{references}
\end{document}